\documentclass[letter, 11pt]{article}
\usepackage{graphicx} 
\graphicspath{{./}{figures/}{../}{../figures/}}

\usepackage[margin=1in]{geometry}
\usepackage{amsmath, amssymb}
\usepackage{mathrsfs}
\usepackage{mathtools}
\usepackage{setspace}
\usepackage[nocompress]{cite}
\usepackage{subcaption}
\usepackage{xcolor}

\newcommand{\dd}{\,\mathrm{d}}
\usepackage{titling}
\usepackage{hyperref}   
\usepackage{cleveref}
\usepackage{array}

\title{Multi-Illumination Spectral Reflectance Reconstruction:\\
A method for hyperspectral imaging using digital photography}
\author{Robert Heyck, Vincent Gillaspy, Stephen J. Hagen, BingKan Xue}
\date{Department of Physics, University of Florida, Gainesville, FL 32611, USA}

\begin{document}

\maketitle

\begin{abstract}
Quantitative measurement of complex coloration patterns is essential for many biological studies and practical applications. Although color perception depends on the observer's visual system, an objective characterization is possible by measuring the spectral reflectance across the sample surface. Traditional methods for measuring spectral reflectance are either limited in spatial resolution, as with spectrometer probes, or limited in spectral detail, as with digital cameras. We present a method for reconstructing spatially resolved spectral reflectance of samples from multiple images taken under different illuminants with known spectral irradiance. Using an optimization scheme that minimizes pixel-value error while regularizing for the smoothness of the reflectance spectrum, our method generates high-resolution hyperspectral images. We tested our method using both existing spectral reflectance datasets and original multi-illumination images taken in a custom photo-box, demonstrating high accuracy and simple implementation of the method, as well as the ability to infer how samples are perceived by different visual systems. Our method can be a useful tool for objective quantification of complex visual patterns in color-related biological research and imaging applications.
\end{abstract}

\section{Introduction}

Quantitative color measurement is important in many fields, including medical imaging \cite{HSI_medical_imaging_dental}, agriculture \cite{HSI_agriculture_ref}, machine vision \cite{machine_vision_color_constancy}, and industrial quality control \cite{textiles_industrial_color_quality_control_review}. In biological research, quantifying animal coloration patterns is crucial for studying species interactions and evolutionary processes, including camouflage, mimicry, aposematism, and sexual selection \cite{Cuthill, Stevens, Cuthill2005Disruptive, sexual_selection_endler, aposematism_and_sexual_selection, mimicry_butterfly_Hoyal-Cuthill}. However, the perception of color is inherently observer-dependent. The light signal is projected onto the observer's visual sensitivity functions to produce the color stimulus values. Two different light spectra can give rise to the same perceived color for a given observer, a phenomenon known as metamerism. Conversely, the same light spectrum can elicit different color perceptions in observers with distinct visual sensitivities.

This distinction is particularly important in biology, where both coloration and color vision are highly diverse among species. Color vision in animals is enabled by the retinal cone cells, which vary widely across species in the number of cone types and their  wavelength sensitivities \cite{Arikawa, Bybee, Cuthill, Stevens}. Humans are trichromats with three types of cones (L, M, S), roughly corresponding to red (R), green (G), and blue (B) wavelengths, covering the so-called visible range of about 400--700~nm \cite{Rowlands, Stevens}. In contrast, most birds are tetrachromats with an additional cone that is sensitive to ultraviolet (UV) wavelengths of about 300 to 400~nm \cite{Bybee, Dell'Aglio, Stevens}. Many insects that are prey for birds, such as butterflies, exhibit UV-reflective patterns that are not visible to humans \cite{Bybee, Dell'Aglio}; even within the visible range, humans and birds perceive different colors owing to differences in cone sensitivities.

To avoid the issue of metamerism, as well as to account for different visual sensitivities, it is important to quantify the visual characteristics of an object using an observer-independent representation. The spectral reflectance of the object's surface, which is the fraction of light it reflects at each wavelength, is a physical property that is independent of the observer \cite{Stevens}. Given the spectral reflectance of the object and the irradiance spectrum of the light source, one can predict how the object will appear to an observer with known visual sensitivities. As the spectral sensitivities of animal photoreceptors have been characterized across a wide range of taxa \cite{OsorioVorobyev2005PhotoreceptorReview, Hart2000PasserineBirdVision, McCulloch2016SexualDimorphismHeliconius, vanDerKooi2021_insect_photoreceptor_review}, inferring how objects will be viewed by different species under specific illumination conditions relies primarily on measuring the spectral reflectance of the surface.

However, biological coloration patterns are often spatially complex, as exemplified by the intricate wing patterns of butterflies. Characterizing such complex coloration patterns requires obtaining reflectance data with both high spectral and high spatial resolution. Traditional spectrometer probes offer high spectral resolution but have little spatial resolution, as they collect light by pointing a fiber optic probe at a single spot on the sample \cite{Bybee, Stevens}. On the other hand, digital cameras provide high spatial resolution, but collect light signals in only a small number of color channels (RGB or monochrome) \cite{chan2022multispectral, Rowlands, Stevens}. The ideal measurement is a so-called hyperspectral reflectance image \cite{Zhang2022_review_reflectance_recon}, which measures a reflectance spectrum at every pixel.

Different imaging techniques have been devised to address these challenges \cite{chan2022multispectral, Shen, Troscianko-Stevens, Valero, Yao, Zhang2022_review_reflectance_recon}. Digital cameras can be furnished with color filters to image across multiple wavelength bands (e.g., R, G, B, UV), producing the so-called multispectral images \cite{chan2022multispectral, Valero, MSIvsHSI_and_reflectanceHSI}. However, the number of wavelength bands isolated by different filters (usually 3--10 bands) falls short of true hyperspectral imaging (on the order of 100 bands) \cite{MSIvsHSI_and_reflectanceHSI}. 
There are commercially available hyperspectral cameras that take hyperspectral images using different techniques, such as line-scanning, wavelength-scanning, and light-field imaging \cite{hyperspectral_snapshot_review, snapshot, snapshot_review_2025, Hong2026}. Despite their high cost, most commercial hyperspectral cameras lack sensitivity in the UV range \cite{Hong2026, Troscianko2025HOSI}. They also involve various architecture-dependent trade-offs among sensitivity, spatial resolution, spectral resolution, dynamic range, and acquisition speed \cite{snapshot_review_2025, hyperspectral_snapshot_review, Troscianko2025HOSI}.

On the other hand, computational methods have been developed to infer hyperspectral images from standard RGB images \cite{Burns, vanTrigt1, vanTrigt2, Yao, Valero, Zhang2022_review_reflectance_recon, sparse_recovery_arad_ben-shahar, dictionary_prior_Fu, local_linear_embeddding_Li, gaussian_process_Akhtar_Mian, HSCNN, HSCNN+, U-Net_Stiebel, adversarial_network_Alvarez-Gila}. These range from simple optimization schemes \cite{Burns, vanTrigt1, vanTrigt2, Yao, Valero} to more sophisticated prior-based methods \cite{Zhang2022_review_reflectance_recon, sparse_recovery_arad_ben-shahar, dictionary_prior_Fu, local_linear_embeddding_Li, gaussian_process_Akhtar_Mian}, as well as data-driven machine learning approaches \cite{HSCNN, HSCNN+, U-Net_Stiebel, adversarial_network_Alvarez-Gila}. However, these techniques may suffer from limitations in spectral accuracy, generality, or applicability to real RGB image data \cite{Zhang2022_review_reflectance_recon}. For example, data-driven machine learning algorithms require extensive training data and may generalize poorly to images in categories outside the training set, while the accuracy of prior-based approaches can depend strongly on the assumptions imposed by the chosen priors \cite{Zhang2022_review_reflectance_recon}.

In this work, we present a method for reconstructing the spectral reflectance of an object from digital images taken under different illumination sources. We tested the method on images taken with a standard RGB camera under multiple LED lights. The use of multiple illuminants with different irradiance spectra circumvents the problem of metamerism and greatly improves the accuracy of spectral reflectance reconstruction. Robust reconstruction is further enhanced by leveraging the smoothness of natural reflectance spectra. Our method is based on the biophysical process of color sensing, in contrast to statistical inference as in machine learning approaches. In the following, we validate our method using two existing spectral reflectance datasets and demonstrate its application on sample images captured in a custom-built multi-illumination photo-box. Our method significantly expands the potential for using digital photography to collect objective and quantitative image data for research and applications.

\section{Methods}

\subsection{Theory and algorithm}

The perception of an object's color depends on three factors: the spectral irradiance of the light source, the spectral reflectance of the object, and the sensitivity functions of the observer's visual system. The color stimulus values registered by the visual system (e.g., cone catch values or camera pixel values) are determined by the equation:
\begin{equation} \label{eq:predict}
\hat{P}_{i}^{(k)}[\rho(\lambda)] = \int S^{(k)}(\lambda) \, c_{i}(\lambda) \, \rho(\lambda) \dd\lambda \,, 
\end{equation}
where $\lambda$ is the wavelength and the integration is over the sensitive range of the visual system \cite{Rowlands, Burns, Zhang2022_review_reflectance_recon, Ribes_Schmitt_linear_inverse_problems}. Here, $S^{(k)}(\lambda)$, for $k = 1, \cdots, N$, are different irradiance spectra of $N$ illuminants; $c_i(\lambda)$, for $i = 1, \cdots, M$, are the sensitivity functions for $M$ types of sensors (cone cells for animal vision or color channels for cameras); $\rho(\lambda)$ is the spectral reflectance of the sample; and $\hat{P}_i^{(k)}$ is the predicted color stimulus value of the $i$-th sensor measured under the $k$-th illuminant. This equation assumes that the sample is not fluorescent, and its reflectance falls within the range $0 \leq \rho \leq 1$. Note that the values of $\hat{P}_i^{(k)}$, $S^{(k)}$, and $c_i$ are subject to arbitrary choices of normalization, which will be discussed in \Cref{sec:pix-norm-scale-factor}.

Thus, the color stimulus values can be thought of as low-dimensional vectors obtained by projecting an infinite-dimensional spectral reflectance curve (weighted by the illuminant spectrum) onto the visual system's sensitivity functions. This is the origin of metamerism, because many different spectral reflectance curves can produce the same color stimulus values for a given visual system under a specific illumination. In other words, the low-dimensional color stimulus values do not uniquely determine the object's spectral reflectance. However, we can reduce the degeneracy by increasing the number of illuminants, since two reflectance curves that are metameric under one illuminant will generically produce distinct color stimulus values under a different illuminant. Therefore, the main idea of our method is to use multiple illuminants with different irradiance spectra to infer the spectral reflectance from many color stimulus values.

\subsubsection{Reconstruction of spectral reflectance}

Our goal is to reconstruct the spectral reflectance $\rho(\lambda)$ from a set of measured color stimulus values $P^{(k)}_i$, given the corresponding illuminant spectra $S^{(k)}(\lambda)$ and sensor sensitivity functions $c_{i}(\lambda)$. Even with multiple illuminants, this inverse problem can be ill-posed because the reflectance spectrum $\rho(\lambda)$ is a continuous function that has (infinitely) many degrees of freedom \cite{Ribes_Schmitt_linear_inverse_problems}. To regularize the problem, we impose the condition that the reflectance curve is smooth, which is true for most natural samples \cite{Burns, vanTrigt1, Troscianko-Stevens}. This can be achieved by minimizing a roughness measure defined as the squared $L^2$ norm of the derivative:
\begin{equation} \label{eq:smooth}
\mathcal{L}_\textrm{r} [\rho(\lambda)] = \frac{1}{2} \int \biggl( \frac{\dd\rho}{\dd\lambda} \biggr)^2 \dd\lambda \,.
\end{equation}
$\mathcal{L}_\mathrm{r}$ has been used as an objective function for $\rho(\lambda)$ \cite{Burns, vanTrigt1, vanTrigt2}, to be minimized under $M$ equality constraints that $\hat{P}_i$ (for a single illuminant) predicted by \Cref{eq:predict} must match the measured pixel values $P_i$. However, such zero-error constraints make this method sensitive to noise in the pixel values and thus prone to overfitting the data.

We therefore propose an objective function that includes the mean squared error (MSE) of the pixel values under multiple illuminations, as well as the roughness of the reflectance curve. With $N$ total illuminants, the objective function is:
\begin{equation} \label{eq:objective}
\mathcal{L}[\rho] = \frac{1}{2} \sum_{k=1}^N \sum_{i=1}^M \Big| \hat{P}^{(k)}_i[\rho] - P^{(k)}_i \Big|^2 + w \, \mathcal{L}_\textrm{r} [\rho] \,.
\end{equation}
Here $\mathcal{L}_\textrm{r}$ serves as a regularization term, with $w$ as a hyperparameter (or ``regularization parameter''). A suitable value for $w$ can prevent the reflectance curve from having drastic oscillations due to overfitting. We will discuss the choice of the parameter $w$ in examples considered below.

To proceed, we discretize the wavelength range into $L$ bins, so that \Cref{eq:predict} takes the form:
\begin{equation} \label{eq:pred-mat}
\hat{\mathbf{P}} = \mathbf{C} \boldsymbol{\rho} \,,
\end{equation}
where $\hat{\mathbf{P}}$ is an $N M \times 1$ vector of predicted color stimulus values (by concatenating all sensor channels and illuminants), $\mathbf{C}$ is an $N M \times L$ matrix constructed from the measured illuminant spectra $S^{(k)}(\lambda)$ and the sensor sensitivity functions $c_i(\lambda)$, and $\boldsymbol{\rho}$ is an $L \times 1$ vector of spectral reflectance. Note that, to approximate an integral over wavelength, $\mathbf{C}$ includes a factor of the wavelength bin-width $\Delta\lambda$. The objective function \eqref{eq:objective} then becomes:
\begin{equation} \label{eq:obj-mat}
    \mathcal{L} = \frac{1}{2} \, (\mathbf{C} \boldsymbol{\rho} - \mathbf{P})^\top \, (\mathbf{C} \boldsymbol{\rho} - \mathbf{P}) + w \cdot \frac{1}{2} \, \boldsymbol{\rho}^\top \, \mathbf{D} \, \boldsymbol{\rho} \,,
\end{equation}
where $\mathbf{D}$ is proportional to the $L \times L$ Laplacian matrix:
\begin{align*}
\mathbf{D} =
\frac{1}{\Delta\lambda}
\begin{bmatrix}
1  & -1 &  &  &  \\[6pt]
-1 & 2  & -1 &  &  \\
& -1 & \ddots & \ddots &  \\
&  & \ddots & 2 & -1 \\[4pt]
&  &  & -1 & 1
\end{bmatrix} ,
\end{align*}
and $\mathbf{P}$ is an $N M \times 1$ vector of the measured color stimulus values. This objective function can be considered a variant of Tikhonov regularization, using the $L^2$ norm of the first derivative rather than that of $\rho$ itself (i.e., using the Laplacian instead of the identity matrix) \cite{Tikhonov-source, Ribes_Schmitt_linear_inverse_problems}.

To find a solution $\boldsymbol{\rho}^*$ that minimizes $\mathcal{L}$, we first ignore the bounds on $\rho$ and set $\frac{\partial \mathcal{L}}{\partial \boldsymbol{\rho}} = 0$, which leads to:
\begin{equation} \label{eq:sol}
\boldsymbol{\rho}^* = (\mathbf{C}^\top \mathbf{C} + w \, \mathbf{D})^{-1} \, \mathbf{C}^\top \, \mathbf{P} \,.
\end{equation}
Note that, when regularization is turned off ($w=0$), this yields the Moore-Penrose pseudoinverse solution to \Cref{eq:pred-mat}. In practice, if the solution \eqref{eq:sol} violates the bounds $0 \leq \rho \leq 1$, we turn to numerical methods for finding $\boldsymbol{\rho}^*$ using SciPy's L-BFGS-B minimization algorithm.
 
Finally, to reconstruct spatially resolved spectral reflectance of a sample, we apply the optimization scheme to each pixel of the sample images. These images should be taken under multiple known illuminants and be perfectly aligned, so that each pixel in one image corresponds to the same pixel in another image. This can be done by switching the illuminants while keeping the camera and sample fixed, as described in \Cref{sec:exp-methods} below. Our optimization algorithm then generates a reflectance spectrum for every pixel, forming a hyperspectral image of the sample.

\subsubsection{Normalization of pixel values and spectral measurements} \label{sec:pix-norm-scale-factor}

\Cref{eq:predict} does not constrain the overall scales of the color stimulus values, illuminant spectra, and sensor sensitivities. The multi-illumination images we used for the spectral reflectance reconstruction have 14-bit pixel values (0--16383). We normalized the pixel values for each color channel, dividing the original pixel values by the pixel value of a Spectralon white standard illuminated by the white LED. This normalization places the pixel values largely below 1; a pixel value can occasionally exceed 1, such as when a blue surface is imaged under a blue LED that is brighter than the white LED in that wavelength range. Note that the same channel-specific normalization is applied to images acquired under all LEDs, rather than normalizing for each illuminant separately. This preserves the relative intensities of the different illuminants and consequently gives greater weight in the objective function to measurements with higher signal-to-noise ratio. 

The illuminant irradiance spectra $S^{(k)}(\lambda)$ were measured with a spectrometer probe (see \Cref{sec:illum-profiles}), which outputs values in arbitrary detector-count units. We normalized them by setting the highest peak among all LEDs to 1 while keeping their relative scales (\Cref{fig:figure1}b). The camera sensitivity curves $c_i(\lambda)$, measured in the lab (\Cref{sec:camera-sens}), were normalized so that the maximum sensitivity of each color channel is 1 (\Cref{fig:figure1}d). Because the irradiance spectra, sensitivity curves, and pixel values should satisfy \Cref{eq:predict}, we rescaled the sensitivity curves by fitting a multiplicative scale factor for each color channel. These scale factors were determined by least-squares fitting using pixel values of a Spectralon white standard and two color patches (``Yellow'' and ``Blue'') on a Calibrite ColorChecker imaged under eight LEDs (see \Cref{sec:exp-methods}).

\subsubsection{Evaluation of reconstruction error} \label{sec:error-metrics}

To verify our method's spectral accuracy when applied to both simulated and real image data, we compare against true reflectance that is either known from a dataset or measured directly. We quantify the reconstruction error by the relative root mean square error (RRMSE) \cite{NRMSE_wheat_reflectance, NRMSE_Clausen}: 
\begin{equation} \label{eq:RRMSE}
\text{RRMSE} = \frac{\sqrt{\frac{1}{L}\sum_i (\rho_i - \rho^*_i)^2}}{\overline{\rho}} \, ,
\end{equation}
where $\rho_i$ is the true reflectance at the $i$-th wavelength bin, $\overline{\rho}$ is the mean of the true reflectance over all wavelengths, and $\rho^*_i$ is the reconstructed reflectance. Compared to the standard root mean square error (RMSE), RRMSE ensures that the same absolute error matters less when the overall reflectance is large, facilitating comparison across samples with significantly different reflectance spectra \cite{NRMSE_Clausen}.

However, spectral error metrics like RRMSE can give a misleading sense of perceptual differences. In humans, for instance, a reconstruction that matches the true spectrum everywhere except near the end of the visible range will produce very similar color stimulus values, despite having a large RRMSE. To capture perceptual differences between pairs of colors, we calculate color stimulus values in the CIELAB color space, which is designed to be approximately perceptually uniform for human observers, then use those to calculate the CIEDE2000 color-difference metric $\Delta E_{00}$. In particular, a CIEDE2000 distance of 1.0 between two colors typically indicates a just-noticeable difference (JND) for human observers \cite{Sharma2003DigitalColorImaging, CIEDE2000}.

For the purpose of visualizing color perception by different species, we calculate the cone-catch values (response of each photoreceptor type) at each pixel of a hyperspectral reflectance image using the known sensitivity functions of human \cite{CIE1931CMF, Rowlands}, bird \cite{Govardovskii2000VisualPigmentTemplate, Hart2000PasserineBirdVision, Hart2005OilDropletAbsorption}, and butterfly vision \cite{McCulloch2016SexualDimorphismHeliconius} under daylight (CIE Standard Illuminant D65). To properly display the resulting images on a screen, we transform the color stimulus values to the standard sRGB space \cite{sRGB_reference} as follows. For humans, we first convert LMS cone-catch values to XYZ by the standard linear transformation; XYZ values are then transformed linearly to the ``linear sRGB'' space, and finally mapped nonlinearly to sRGB via the standard transfer function. For bird and butterfly, we perform the same transformations using the bird and butterfly LMS values, with an additional step to normalize the linear sRGB values so that perfect white ($100\%$ reflectance) is mapped to $[255, 255, 255]$. The resulting images are meant as a visualization of how the bird's and butterfly's color perceptions differ from humans.

\subsection{Experimental methods} \label{sec:exp-methods}

We tested our spectral reflectance reconstruction method using both synthetic images generated from existing spectral reflectance datasets and real images of color patches and butterfly specimens photographed in the lab. To apply the method, we constructed a custom photo-box with multiple illuminants installed, and used a digital camera that had been modified to allow UV-visible photography. Furthermore, we constructed a wavelength-scanning hyperspectral imaging setup for direct measurement of spectral reflectance, which is used as a reference for verifying and calibrating our reconstruction method.

\subsubsection{Multi-illumination photo-box} \label{sec:photo-box}

\begin{figure}[t]
    \centering  
    \includegraphics[width=\textwidth]{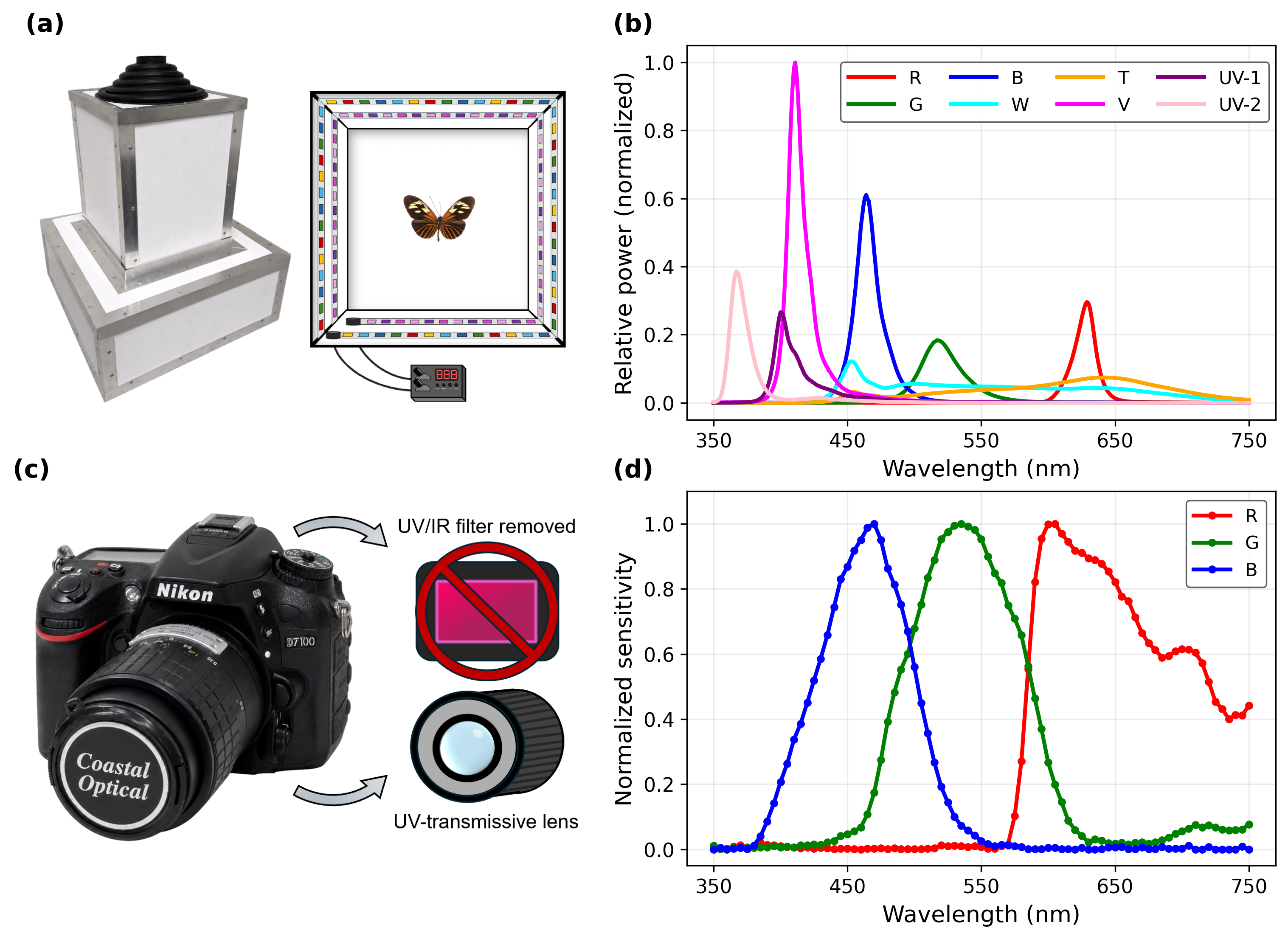}
    \caption{\small Components of the multi-illumination imaging setup. (a) Custom photo-box made for multi-illumination imaging. The left image shows the exterior, and the right schematic plot shows the interior lined with reflective material; LED strip lights are controlled by an external switch. (b) Relative irradiance spectra $S^{(k)}(\lambda)$ of the LED illuminants. The peak of the brightest LED (Violet) is scaled to 1. (c) Digital RGB camera used in multi-illumination imaging, with modifications to allow near-UV and near-IR imaging, paired with a UV-transmissive lens. (d) Relative spectral sensitivity functions $c_i(\lambda)$, sampled every 5~nm, for the RGB camera. The peak sensitivity of each channel is normalized to 1; the standard error (not shown) is less than 0.1\% of the peak.}
    \label{fig:figure1}
\end{figure}

We constructed a custom photo-box (\Cref{fig:figure1}a) to image the samples under controlled illumination in the absence of ambient light. The interior of the lower chamber was lined with highly reflective white acrylic to provide near-uniform diffuse illumination. Inside the lower chamber, we installed eight LED illuminants, referred to as red (R), green (G), blue (B), white (W), tungsten (T), violet (V), UV peaked at 395~nm (UV-1), and UV peaked at 365~nm (UV-2). The first five were part of a 5-in-1 LED strip (Waveform Lighting FilmGrade FiveSpect), while the last three were individual LED strips (Waveform Lighting SimpleColor Violet, realUV 395~nm, and realUV 365~nm). Together, their irradiance spectra span the UV-visible range of approximately 350--750~nm (\Cref{fig:figure1}b). The camera was mounted above an opening at the top of the removable upper chamber; a flexible rubber hood was fitted around the lens to block stray light. Samples were placed at the center of the lower chamber, and images were taken while the LEDs were switched on sequentially. 

Images were captured by a modified digital SLR camera (Nikon D7100), which had its internal UV/IR blocking filter removed (\Cref{fig:figure1}c). The camera was equipped with a UV-transmissive lens (Coastal Optics 60mm 1:4 UV-VIS-IR Apo Macro), which is transmissive and apochromatic within 315--1100~nm. To preserve the linear relationship between pixel values and light intensity, we used a fixed exposure time (1/100~s) that keeps the CMOS sensor responses within the linear regime, and saved all images in RAW format with white balance disabled \cite{Stevens}. To reduce pixel noise, we took multiple images under each illumination condition and took the pixel-wise average over the images. We used a custom procedure to demosaic and downscale the raw images: every block of $4 \times 4$ pixels is merged into one pixel whose RGB values are calculated as the median of the correspondingly colored pixels according to the Bayer filter layout. This procedure yielded RGB images of 1509$\times$1005 pixels, with a field of view of about $16.9 \, \text{cm} \times 11.2 \,\text{cm}$.

\subsubsection{Illumination profiles} \label{sec:illum-profiles}

To measure the spectral irradiance $S^{(k)}(\lambda)$ of our LED illuminants (\Cref{fig:figure1}b), we used a spectrometer probe (Labsphere CDS-600) and a nominal 99\% white diffuse reflectance standard (Spectralon SRS-99-010), which is spectrally flat across UV-VIS-NIR and designed for use over the wavelength range of 250--2500~nm \cite{Labsphere2019Spectralon}. We placed the white standard inside the photo-box at the same height as the samples to be imaged, positioned the spectrometer probe at a fixed distance above it, and then collected (uncalibrated) spectral radiance data as the LEDs were switched on sequentially. The accuracy of the spectrometer's wavelength readings were verified using a holmium oxide wavelength calibration standard (Spectralon WCS-HO-010), which has absorbance peaks at known wavelengths. The spectrometer's readout (in arbitrary detector-count units) was calibrated using a traceable tungsten-halogen lamp (Labsphere SCL-600 Spectral Flux Standard) inside an integrating sphere (Labsphere illumiaPlus). This calibration allowed us to correct for the wavelength-dependent instrument response in our radiance measurements and, assuming Lambertian 100\% reflection from the white standard, obtain the relative spectral irradiance of each LED.

Although the photo-box interior was lined with reflective material, the imaging area within the photo-box was not uniformly illuminated by the LEDs. To correct for the spatial variation, we measured the LED spectra with the white standard placed at multiple locations within the sample imaging area (about $5.8 \, \text{cm} \times 8.7 \,\text{cm}$). We found that, while the shape of the irradiance spectra remained the same for each LED, their overall intensity increased by roughly 10\% from the center of the photo-box to the edges of the sample imaging area. To account for such variation, we fitted the spatial dependence of light intensity with a second-degree polynomial function. The illuminant irradiance spectra were therefore scaled by this spatially dependent intensity at each pixel.

\subsubsection{Hyperspectral measurements} \label{sec:hyperspectral-camera}

We constructed a wavelength-scanning hyperspectral imaging apparatus (Supplementary \Cref{fig:hyperspectral-setup}), which was used to validate our spectral reflectance reconstruction method and to obtain the camera sensitivity functions $c_i(\lambda)$ of our digital camera (see next section). Light from a short-arc Xenon lamp (Perkin Elmer LS-1130-1) was focused by a fused silica lens into the entrance slit of a grating monochromator (Acton SpectraPro-300i). Monochromatic light was collected from the exit slit and delivered into an optical enclosure (Thorlabs XE25C9) using a multimode optical fiber. The light was then passed through a UV fused silica engineered diffuser (Thorlabs EDG5C20M) to illuminate the sample object. The sample was imaged using a scientific monochrome CCD camera (Andor Clara). Meanwhile, a spectrometer probe (LabSphere CDS-600) was placed near the sample to measure the reflected light.

The monochromator's wavelength output was verified by using the spectrometer probe to measure the light scattered from a white standard. The output monochromatic light had a bandwidth of approximately 13.4~nm. The monochromator was controlled by a computer to scan wavelengths from 350 to 750~nm in 5~nm steps. Images of the sample were taken at each step and assembled into a hyperspectral image. The reflectance of the sample at each selected wavelength and each pixel was calculated by taking the ratio of the sample image to that of a large Spectralon white plate. The images were 1000$\times$1000 pixels, with an imaging area of $69.4 \, \text{mm} \times 69.4\,\text{mm}$. The camera exposure time was chosen to maximize signal while keeping the raw pixel values within the linear range (below 37\% of the full 16-bit scale).

\subsubsection{Camera sensitivities} \label{sec:camera-sens}

The spectral sensitivity functions $c_i(\lambda)$ of the digital camera were measured using the hyperspectral imaging setup as follows. The camera was placed inside the optical enclosure (Supplementary \Cref{fig:hyperspectral-setup}) and used to image a white standard illuminated by monochromatic light at each selected wavelength. Simultaneously, the spectrometer probe was used to measure the reflected light from the white standard. For each selected wavelength, the total irradiance was estimated by integrating the measured spectrum over wavelength. The color stimulus values of the RGB image, averaged over an illuminated area of $64 \times 100$ pixels, were then divided by the total irradiance to yield the camera's spectral sensitivity in each color channel (\Cref{fig:figure1}d). We found that the camera sensitivity dropped to zero below about 370~nm, while the irradiance of the xenon lamp dropped off above 720~nm. Therefore, we restricted our spectral reflectance reconstructions to 370--720~nm.

\subsubsection{Test datasets and samples}

To test our method of spectral reflectance reconstruction, we evaluated it on four datasets, including two publicly available datasets of reflectance spectra and hyperspectral images, and two custom datasets consisting of images and spectrometer measurements collected in the lab.

\begin{itemize}
\item The Munsell dataset \cite{CentoreMunsellGlossyReflectances} consists of spectral reflectance curves of 1484 paints from X-Rite's 2007 Munsell Book of Color, which is used as an industry standard for color matching. These Munsell colors have diverse, smooth reflectance curves in the visible range of 400--700~nm, with a resolution of 31 wavelength bands at 10~nm intervals.
\item The CAVE dataset \cite{CAVE_dataset_report} consists of multispectral images of 32 diverse scenes. These images have a spatial resolution of $512 \times 512$ pixels, and a spectral resolution of 31 wavelength bands in the 400--700~nm range.
\item A ColorChecker card (Calibrite Classic Mini $63.5 \times 109$~mm) contains 24 uniform color patches, often used in photography as a reference for adjusting white balance and color profiles. We measured the spectral reflectance curves of these color patches using both the spectrometer probe and the hyperspectral imaging setup.
\item Four butterfly specimens were provided by the McGuire Center for Lepidoptera and Biodiversity at the Florida Museum of Natural History. There were two mimetic species pairs within the \textit{Heliconius} genus: \textit{H. erato petiverana} and \textit{H. melpomene rosina}, \textit{H. erato amazona} and \textit{H. melpomene meriana}. We imaged their dorsal wings using both the photo-box and the hyperspectral imaging setup. In addition, we used the spectrometer probe to measure a few distinct color patches on the wing that are shared by the species.
\end{itemize}

\section{Results} \label{sec:results}

We applied our spectral reflectance reconstruction method to the published datasets and measured samples as listed in \Cref{tab:reconstruction-tests}. We used the published datasets to test the accuracy and consistency of our method. These reflectance data served as ground truth, from which we generated synthetic color stimulus values using given illuminant spectra and sensor sensitivities. We then tested whether our method could recover the original reflectance from these hypothetical noise-free sensor responses. Furthermore, we used the measured samples to test the accuracy and robustness of our method when applied to real image data with potential measurement noise. We tested our method on datasets of both single colors, including the Munsell colors and the ColorChecker, and spatially complex patterns, including scenes from the CAVE dataset and butterfly images taken in the lab.

\begin{table}[t]
\centering
\caption{Four tests used to evaluate the multi-illumination spectral reflectance reconstruction.}
\label{tab:reconstruction-tests}
\small
\renewcommand{\arraystretch}{1.55}
\setlength{\tabcolsep}{6pt}
\begin{tabular}{>{\raggedright\arraybackslash}p{0.12\textwidth}
>{\raggedright\arraybackslash}p{0.24\textwidth}
>{\raggedright\arraybackslash}p{0.31\textwidth}
>{\raggedright\arraybackslash}p{0.23\textwidth}}
\hline
Samples & Inputs & Outputs & Metric \\
\hline
Munsell colors \cite{CentoreMunsellGlossyReflectances}
& Published reflectances, 6 LED irradiances, XYZ color matching functions
& Single-color reflectance spectra reconstructed from calculated XYZ tristimulus values
& Reflectance RRMSE against published values \\

ColorChecker
& Measured RGB images, 8 LED irradiances, camera sensitivities
& Single-color reflectance spectra reconstructed from measured RGB pixel values
& Reflectance RRMSE \& CIEDE2000 color differences against measured values \\

CAVE dataset \cite{CAVE_dataset_report}
& Published hyperspectral reflectance images, 6 LED irradiances, XYZ color matching functions
& Hyperspectral reflectance images reconstructed from calculated XYZ tristimulus values
& CIEDE2000 color differences for each pixel \\

Butterfly specimens
& Measured RGB images, 8 LED irradiances, camera sensitivities
& Hyperspectral reflectance images reconstructed from measured RGB pixel values
& Reflectance RRMSE for selected color patches \\
\hline
\end{tabular}
\end{table}

\subsection{Single color reconstruction}

To verify that our method works for a wide range of colors, we first applied it to the Munsell dataset, which provides a broad and perceptually uniform sample of the human color gamut \cite{newhall1943final}. We then applied our method to real images of a Calibrite ColorChecker card captured in the photo-box.

\subsubsection{Munsell colors}
\label{sec:munsell} 

To generate synthetic color stimulus values for each of the 1484 Munsell colors, we first chose a set of illuminant spectra $S^{(k)}(\lambda)$ and a set of sensor sensitivities $c_i(\lambda)$. The sensitivity functions were chosen to be those of the CIE 1931 $2^\circ$ standard colorimetric (human) observer, given by the CIE XYZ color-matching functions $\bar{x}(\lambda), \bar{y}(\lambda), \bar{z}(\lambda)$ \cite{CIE1931CMF}. The illuminant spectra were chosen to be those of our LEDs; because the Munsell reflectance data only span the wavelength range of 400--700~nm, we used the six non-UV LEDs. The XYZ tristimulus values of each Munsell color under each illuminant were then calculated from \Cref{eq:predict} using the spectral reflectance data $\rho(\lambda)$ from the dataset. 

\begin{figure}[t]
    \centering  
    \includegraphics[width=\textwidth]{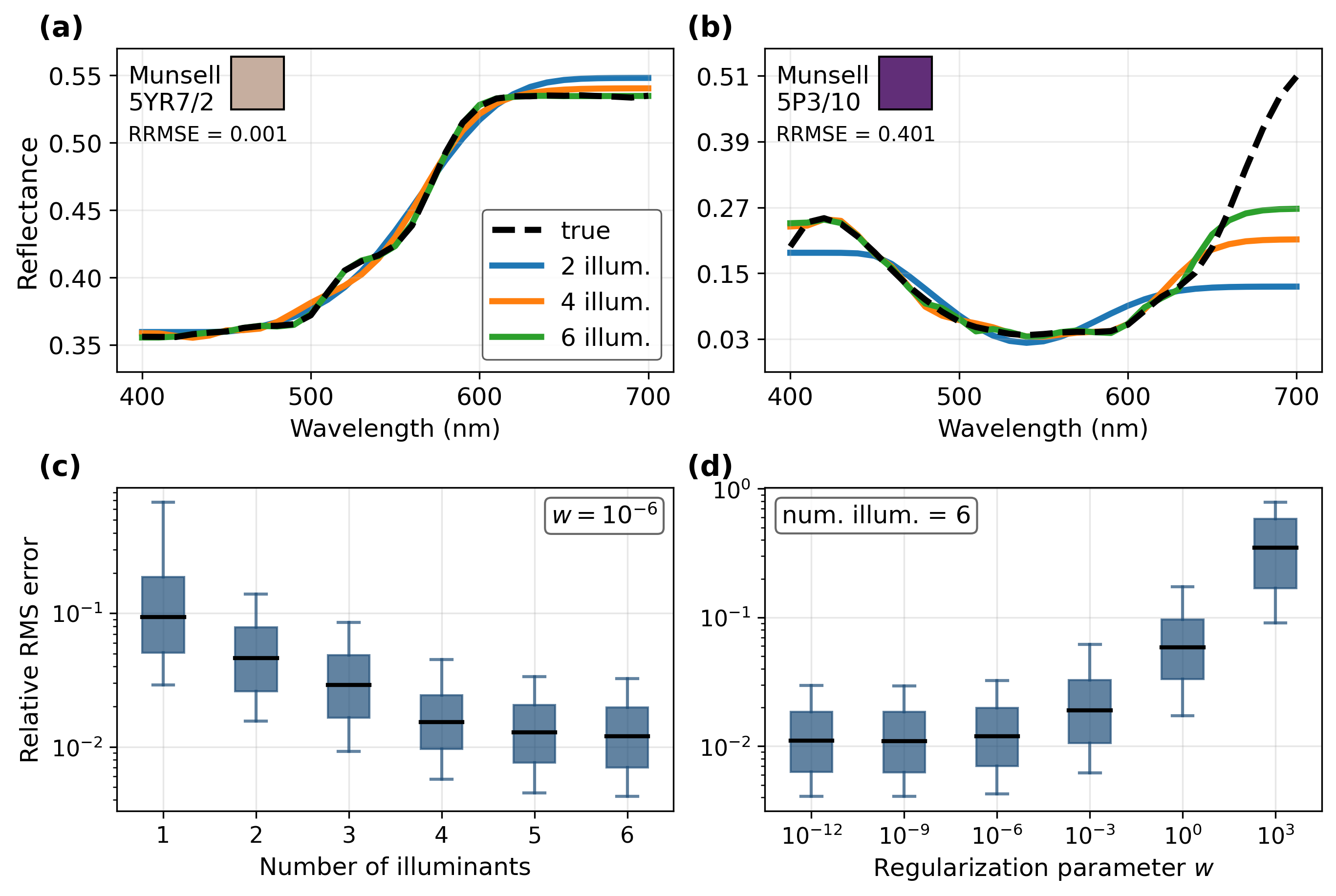}
    \caption{\small Spectral reflectance reconstruction results for the Munsell dataset. (a, b) Examples of reconstructed reflectance curves using different numbers of illuminants (solid curves). The relative root mean square error (RRMSE) is calculated between the true reflectance (dashed black curves) and the six-illuminant reconstruction (green curves). (c) RRMSE for the whole Munsell dataset (1484 samples) plotted against the number of illuminants used in the reconstruction. The order of the illuminants used is W, T, V, G, R, B. (d) RRMSE plotted against the regularization parameter $w$ when all six illuminants are used. The parameter value $w = 10^{-6}$ is used in (a)--(c). In (c) and (d), black lines indicate the median, blue boxes span the interquartile range, and the whiskers extend from the 10th to the 90th percentile.}
    \label{fig:Munsell}
\end{figure}

For each Munsell color, we applied our reconstruction method using a wide range of regularization parameter values. We also varied the number of illuminants used in the reconstruction from one to six, in the order of W, T, V, G, R, B (with the broadband LEDs first so that the full wavelength range is covered). \Cref{fig:Munsell} summarizes the results of these reconstructions. In most cases (e.g., \Cref{fig:Munsell}a), only 2--4 illuminants are needed to reconstruct an accurate spectral reflectance curve. In some rare cases (e.g., \Cref{fig:Munsell}b), using up to 6 illuminants still produces noticeable error near the edge of the wavelength range. This occurs mainly for colors like violet and red, whose reflectance curves rise steeply at those wavelengths, where low illumination leads to weak constraints. Even in those cases, the reconstructed reflectance curves still match the true curves quite well over most of the wavelength range.

\Cref{fig:Munsell}c quantifies the overall reconstruction error (RRMSE) over the whole dataset as a function of the number of illuminants used. The reconstruction error decreases as more illuminants are used, but levels off with around 4--5 illuminants. The value of the regularization parameter was found to not impact this trend. Furthermore, this trend does not depend much on the choice of the illuminants as long as they adequately cover the wavelength range; indeed, we obtained similar results with six broadband (blackbody) illuminants and with six evenly spaced, sufficiently wide Gaussian curves (Supplementary see \Cref{fig:munsell-other-illuminants}).

\Cref{fig:Munsell}d illustrates how the reconstruction accuracy depends on the value of the regularization parameter $w$. In particular, larger values of $w$ increase the error because they make the reflectance curves too flat, while a small $w$ value (e.g., $10^{-6}$) is sufficient to produce very low reconstruction errors. This is largely due to the use of noiseless synthetic data here. Extremely small values of $w$ can work because it is possible to achieve perfect matching of the tristimulus values when the reflectance curve approaches the true curve. However, for real image data with noise, perfect matching will not be possible, and the error is expected to increase due to overfitting when $w$ becomes too small (see below). These results for the Munsell dataset demonstrate that the spectral reflectance can be accurately reconstructed using a small number of illuminants. Furthermore, our method is robust under different choices of illuminants and is accurate over a range of regularization parameter values.

\subsubsection{ColorChecker} \label{sec:color-checker}

\begin{figure}[t]
    \centering
    \includegraphics[width=\textwidth]{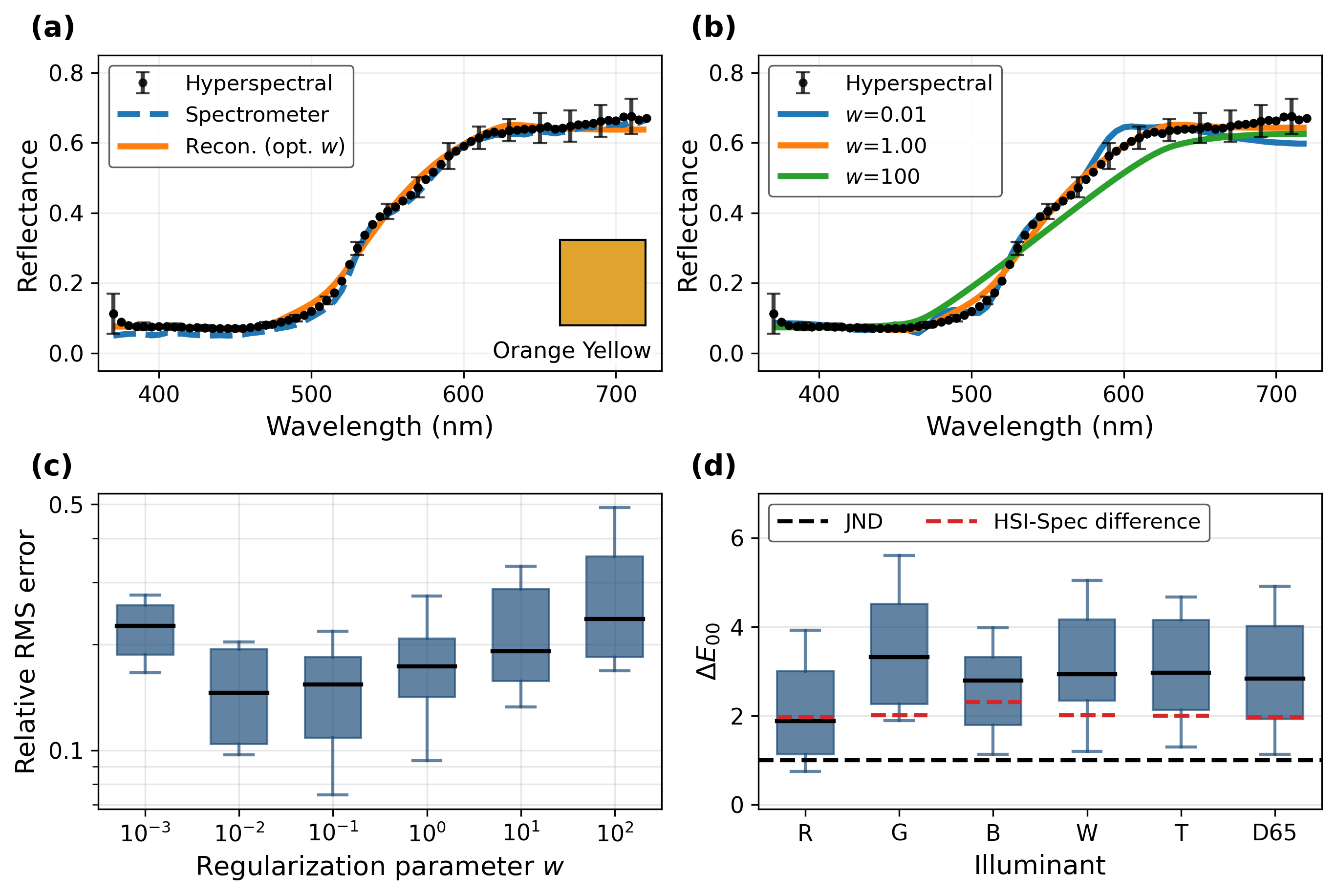}
    \caption{\small Spectral reflectance reconstruction results obtained from images of a Calibrite ColorChecker card. (a) Reconstructed reflectance curve for the ``orange yellow'' color patch (solid orange curve) using an optimal regularization parameter $w=0.59$. Direct measurements of the spectral reflectance are obtained from a hyperspectral imaging setup (black dots) and a spectrometer probe (dashed blue curve). Error bars for the hyperspectral measurement represent the standard deviation over a $10\times10$ pixel region. (b) Comparison of reconstructions using different regularization parameters, for the same color patch as in (a). (c) RRMSE of all 24 color patches as a function of the regularization parameter $w$. Measured reflectance from the hyperspectral imaging setup was used as reference. (d) CIEDE2000 color difference $\Delta E_{00}$ between hyperspectral measurements and reconstructed reflectance for six different illuminants and all 24 colors, where reconstruction was performed with all eight LED illuminants and $w = 0.1$. The dashed black line indicates a just-noticeable difference (JND) according to the standard human observer ($\Delta E_{00} = 1.0$). The dashed red lines indicate the median of the CIEDE2000 color differences between the hyperspectral image (HSI) and the spectrometer (Spec) measurements for all 24 patches.}
    \label{fig:color-checker-summary}
\end{figure}

We next tested our method on images of a Calibrite ColorChecker card with 24 color patches, each about $12 \times 12$~mm in size. We captured digital images of the card using the photo-box under each of the eight LEDs. The images were demosaiced and downscaled as described in \Cref{sec:photo-box}, and the color stimulus values were normalized as described in \Cref{sec:pix-norm-scale-factor}. Each color patch occupies about 100 pixels in width (after downscaling), with small variations in color stimulus values across the patch. Spatial variation in the light intensity was accounted for using the illuminant profiles, as described in \Cref{sec:illum-profiles}. For each color patch, color stimulus values were averaged over a $10 \times 10$ pixel region extracted from the processed RGB images. The resulting mean color stimulus values were used to reconstruct the spectral reflectance of each patch over the wavelength range of 370--720~nm at 5~nm intervals. \Cref{fig:color-checker-summary}a shows an example of the reconstructed reflectance curve, using a regularization parameter of $w = 0.59$ (which is optimal according to an adaptive method described in \Cref{sec:choosing-hyperparameter}).

In order to evaluate the accuracy of the reconstructions, we directly measured the spectral reflectance of these color patches using both the spectrometer probe and our hyperspectral imaging setup. For the spectrometer measurements, we put the ColorChecker in the photo-box illuminated by a combination of the W, UV-1, and UV-2 LEDs, and pointed the probe at each color patch from a close, fixed distance. Since the spectrometer probe collects light from within a small angle, the result represents an average over a small region within a color patch. To obtain reflectance, we took a similar measurement of the white standard and took the ratio of each color patch to the white standard. For the hyperspectral measurements, we took hyperspectral images of the ColorChecker card (\Cref{sec:hyperspectral-camera}) and averaged over a $10 \times 10$ pixel region within each color patch. These different measurements agreed well with each other, as shown by an example color patch in \Cref{fig:color-checker-summary}a (see Supplementary \Cref{fig:color-checker} for all color patches).

\Cref{fig:color-checker-summary}b shows the reconstruction for the same color patch using a variety of regularization parameter values. Among these values, $w=1.0$ appears to give the best overall reconstruction in terms of accuracy and stability, and this accords well with the optimal parameter value used in \Cref{fig:color-checker-summary}a. \Cref{fig:color-checker-summary}c shows how RRMSE varies with the parameter $w$ across all 24 ColorChecker patches, using the hyperspectral measurements as the true values. Both too small and too large parameter values lead to poor reconstructions. For small values of $w$, reconstructions tend to overfit to the noise, leading to spurious oscillations and high variance across similar pixels, whereas large $w$ values prioritize smoothness over agreement with measured color stimulus values, resulting in excessive flattening of the reconstructed spectrum. These results indicate that noisy image data require more careful choices of the regularization parameter, which generally needs to be larger than for noiseless synthetic data (cf. \Cref{fig:Munsell}d).

\Cref{fig:color-checker-summary}d shows color differences according to the CIEDE2000 metric (\Cref{sec:error-metrics}) between the reconstructed reflectance and the hyperspectral measurements of the 24 color patches under six different illuminants (five of our LEDs plus D65). Median distances lie within roughly 2--3 JND, indicating that the reconstructed spectra produce colors that are barely distinguishable from the measured spectra to a human observer \cite{CIEDE2000_evaluation}. We note that the spectra measured by the spectrometer and by hyperspectral imaging also differ by roughly 2 JND (\Cref{fig:color-checker-summary}d, dashed red lines). Therefore, our reconstruction method achieves an accuracy comparable to these direct measurement methods.

\subsection{Full image reconstruction}

The reconstruction of spectral reflectance can be performed on an entire image by applying our method to individual pixels. As a test, we first applied the method to the CAVE multispectral dataset, which consisted of 32 hyperspectral reflectance images. We used them to generate synthetic RGB images and tested the ability of our method to reconstruct hyperspectral images. Moreover, we mapped both the original and reconstructed hyperspectral images onto the CIELAB color space to check whether the images are distinguishable by an observer. We then tested the method on real images of butterfly specimens taken in our photo-box. To demonstrate the use of our method for integrating animal vision, we mapped the reconstructed hyperspectral images of the butterflies onto the visual systems of avian predators and conspecific butterflies.

\subsubsection{CAVE dataset}

As with the Munsell dataset, we first calculated synthetic color stimulus values by projecting the reflectance data of each pixel in a hyperspectral image onto our six non-UV LED irradiance spectra and the three CIE XYZ sensitivity functions. We then performed reconstruction of the spectral reflectance pixel by pixel to obtain reconstructed hyperspectral images for all 32 scenes in the dataset. To evaluate the accuracy of reconstruction with respect to human color perception, we mapped both the true and reconstructed hyperspectral images to the CIELAB color space under daylight illuminant D65, then calculated the CIEDE2000 color difference $\Delta E_{00}$. Note that, because CIELAB is a nonlinear transformation of CIE XYZ, we must use the (linear) CIE XYZ color stimulus values, rather than CIELAB, for performing the spectral reflectance reconstruction.

\begin{figure}[t]
    \centering
    \includegraphics[width=\textwidth]{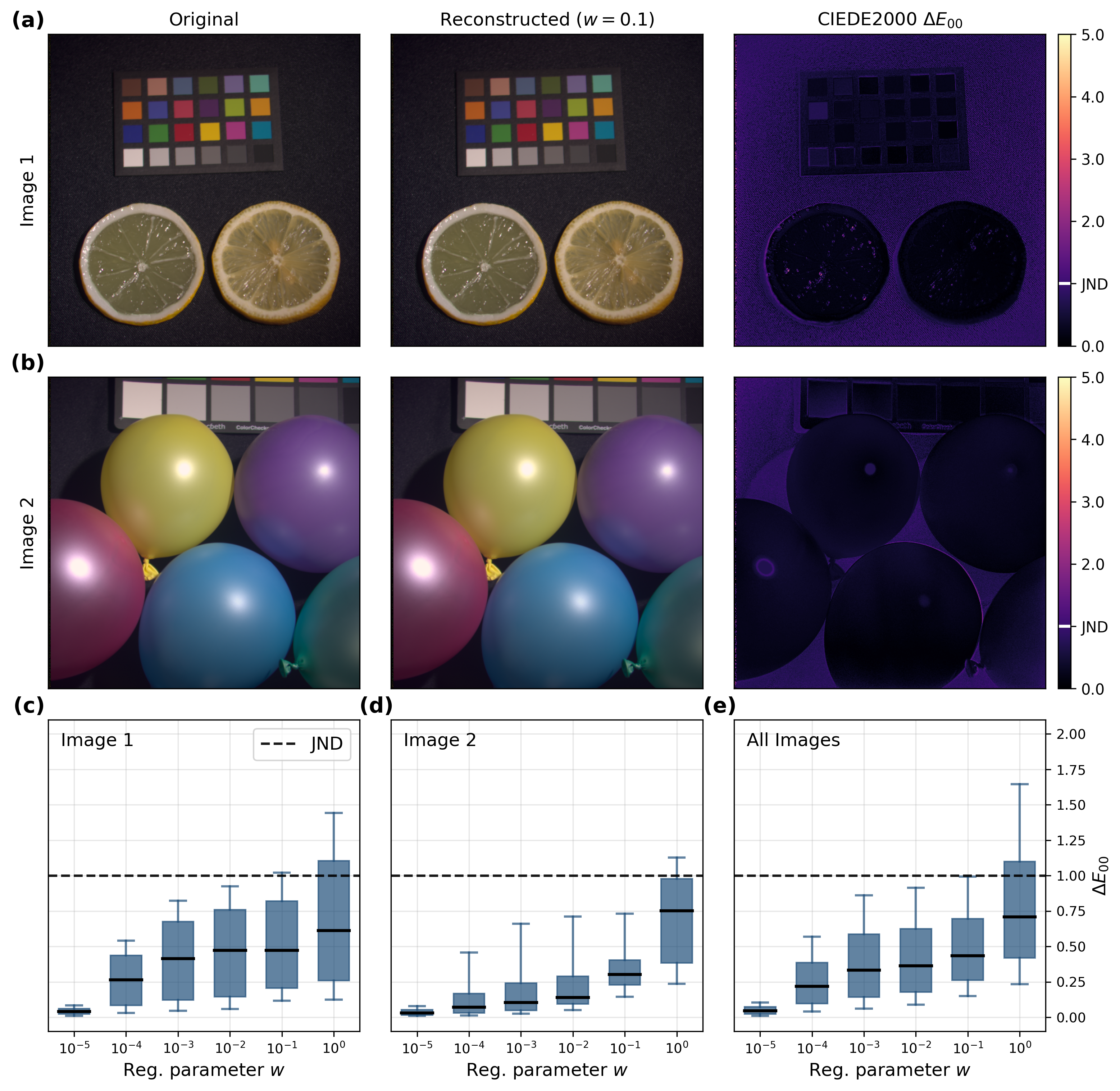}
    \caption{\small Spectral reflectance reconstruction results for the CAVE multispectral image dataset. (a,b) Two scenes from the dataset. Left and center panels show the original and reconstructed (using $w = 0.1$) images projected onto the human visual system under daylight (CIE D65) and converted to the sRGB color space, and right panels show heat maps indicating the per-pixel CIEDE2000 color difference $\Delta E_{00}$ between the two. (c,d,e) Box plots showing the distribution of CIEDE2000 color distances as a function of the regularization parameter $w$, for all pixels in (c) Image 1, (d) Image 2, and (e) the entire dataset (32 images). Dashed black line indicates a just-noticeable difference (JND) according to the standard human observer ($\Delta E_{00} = 1.0$).}
    \label{fig:CAVE}
\end{figure}

\Cref{fig:CAVE}a,b (middle column) show the results of reconstruction for two scenes from the dataset, using a constant regularization parameter $w = 0.1$. Per-pixel CIEDE2000 color differences between the true and reconstructed images are shown in \Cref{fig:CAVE}a,b (right column). It can be seen that the reconstructed images are almost everywhere indistinguishable from the true images as per JND. \Cref{fig:CAVE}c--e shows boxplots of the CIEDE2000 color differences as a function of the parameter $w$ for the two examples shown above, as well as for the entire dataset. As with the Munsell dataset, for synthetic image data without measurement noise in the pixel values, a small parameter $w$ allows accurate spectral reflectance reconstruction.

\subsubsection{Butterfly specimens} \label{sec:butterfly-recon}

\begin{figure}[t]
    \centering
    \includegraphics[width=\textwidth]{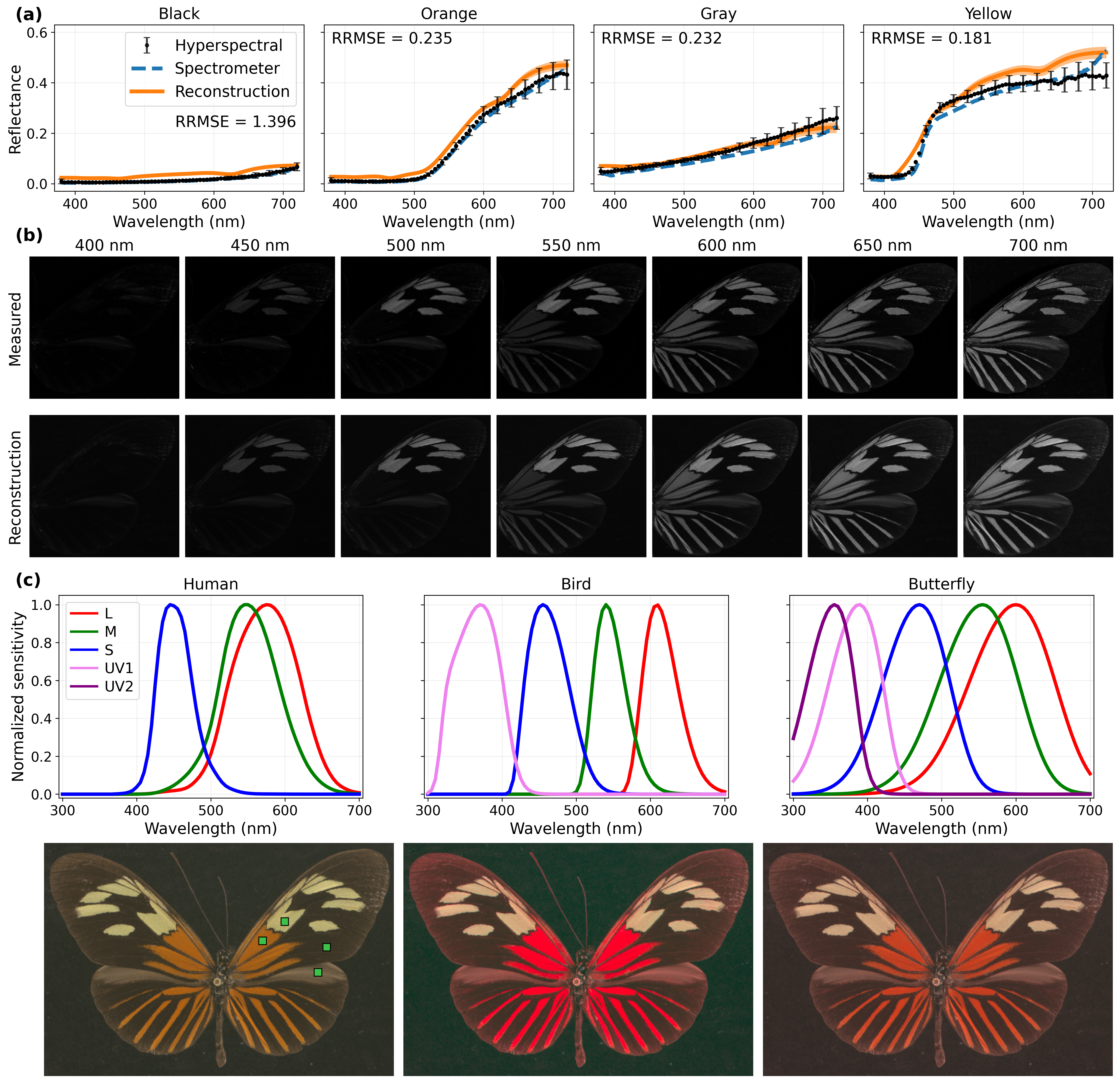}
    \caption{\small Spectral reflectance reconstruction results for a butterfly specimen (\textit{H. erato amazona}) with regularization parameter $w=1$. (a) Reflectance curves for four color patches on the dorsal wings, as indicated in (c) lower left panel. The reconstructed reflectance represents the mean (solid orange curve) and standard deviation (light orange shade) over a $10 \times 10$ pixel region. Direct measurements using the hyperspectral imaging setup (black dots) and spectrometer probe (dashed blue curve) are shown as references. Error bars for the hyperspectral measurement represent standard deviation over a $10\times10$ pixel region. RRMSE is calculated between the reconstructed spectrum and the spectrometer measurement. (b) Comparison of the reconstructed and directly measured hyperspectral images at various wavelengths. (c) Projection of the reconstructed hyperspectral images onto human, bird, and butterfly vision under daylight (CIE D65), converted to the sRGB color space. Upper row shows the cone sensitivities of all three species. Since the sRGB space is three-dimensional, we used only the L, M, and S cones for all three species. The green squares in the lower left panel indicate the four color patches used in (a).}
    \label{fig:b3_combined}
\end{figure}

To test our method on actual images that inevitably have measurement noise in pixel values, we applied the method to images of butterfly specimens we took in the photo-box. The images were processed in the same way as for the ColorChecker, and we ran our algorithm for each pixel. For comparison, we also imaged each butterfly in the hyperspectral imaging setup to obtain the spectral reflectance of the right dorsal wings. In addition, we used the spectrometer probe to measure four spots within the black, orange, gray, and yellow patches on the wings (\Cref{fig:b3_combined}c bottom left panel).

\Cref{fig:b3_combined}a shows the comparison of spectral reflectance curves for one of the butterfly specimens (\textit{H. erato amazona}) across the four spots. We calculated the mean and standard deviation of the reconstructed reflectance curves over a $10 \times 10$ pixel region around each spot. Similarly, the hyperspectral measurements are obtained over a $10 \times 10$ pixel area to calculate the mean and standard deviation. Note that the variance over these spatial regions may reflect both measurement noise and actual biological variation in wing coloration. It can be seen that the reconstructed spectral reflectance agrees well with both the spectrometer and the hyperspectral measurements across the four spots.

For the same butterfly specimen, \Cref{fig:b3_combined}b shows a comparison of the reconstructed and directly measured hyperspectral images at different wavelengths. The match is generally satisfactory, although we could not compare these hyperspectral images pixel by pixel because they were taken separately and not perfectly aligned. The reconstructions are systematically brighter for the yellow patch on the forewing, and to a lesser extent for the orange patch. This discrepancy, which is also seen in the other butterfly specimens (see Supplementary \Cref{fig:other-butterflies}), was found to be caused by these regions of the wings being partially transparent. In the photo-box, the bright illumination and reflective background allowed light to be transmitted from behind the wings, whereas the hyperspectral imaging setup had dimmer, direct illumination and negligible light transmission from behind. Accordingly, the reconstructions and spectrometer measurements, which were both obtained in the photo-box, agreed better with each other (Supplementary \Cref{fig:other-butterflies}).

To demonstrate the usefulness of our method for studying animal vision, we used the reconstructed hyperspectral images to infer how the butterflies look under daylight (CIE D65) according to human, bird (blue tit), and conspecific butterfly (\textit{H. erato}) vision. Human LMS cone responses were calculated from the CIE XYZ sensitivity functions using the standard Hunt-Pointer-Estevez linear transformation \cite{Rowlands}. The bird cone sensitivities were obtained using a well-established template for oil droplet and ocular medium transmittance with peak wavelengths of the blue tit's four cones \cite{Govardovskii2000VisualPigmentTemplate, Hart2000PasserineBirdVision, Hart2005OilDropletAbsorption}. The butterfly cone sensitivities were obtained using the same template with perfect transmittance and previously measured peak wavelengths \cite{McCulloch2016SexualDimorphismHeliconius}. These cone sensitivities are shown in the top row of \Cref{fig:b3_combined}c. 

The bottom row of \Cref{fig:b3_combined}c shows sRGB images created using the LMS cone sensitivities of each species (see \Cref{sec:error-metrics}). Note that, although birds and butterflies have more than three cones, we neglected the UV-sensitive cones, which was partly justified by the fact that the imaged butterfly samples had low UV reflectance. While these images may not truly represent how these species perceive the samples, they clearly illustrate the differences in the color vision of these species. These results thus highlight how our method can be combined with animal color vision models to obtain useful representations of visual patterns from the animal's perspective.

\section{Discussion}

\subsection{Choice of regularization parameter} \label{sec:choosing-hyperparameter}

In testing our method on synthetic images generated from the Munsell and CAVE datasets, reconstruction accuracy remained high across many orders of magnitude of the regularization parameter, including very small values. On the other hand, for experimentally measured ColorChecker and butterfly data, a larger parameter (stronger regularization) was required to prevent overfitting. This is because various sources of measurement noise, as well as any calibration error and non-ideal imaging conditions, can introduce variations in the measured pixel values. In such cases, exact matching of pixel values is not meaningful, making the regularization more important. In practice, we found that a parameter value of the order $w \sim 1$ tends to work well in most situations, provided that the pixel values are normalized to the approximate range of $[0,1]$.

In choosing $w$ the user needs to manage the tradeoff between finding a better fit to the data and avoiding overfitting noisy data. To address this problem without introducing an explicit noise model, we provide an adaptive method that finds an optimal $w$ value for each reconstruction, such that the result is most stable (see Supplementary \Cref{fig:choosing-hyperparameter}). Briefly, consider the reconstructed reflectance spectrum $\rho^*$ as a vector that depends on the parameter $w$. We then look for the value $w = \tan\theta$ that minimizes the squared norm of the sensitivity, $\big| \frac{\partial\rho^*}{\partial\theta} \big|^2$. This minimization has a geometric interpretation: on a plot of pixel-error versus roughness (i.e., the first and second term in \Cref{eq:objective}), $w$ is the slope of the Pareto front, and $\theta$ is the angle of the tangent (Supplementary \Cref{fig:choosing-hyperparameter}). Therefore, the optimal value of $w$ found by this method corresponds to the solution $\rho^*$ that is least sensitive to changes in $\theta$. This method allows the reconstruction algorithm to automatically choose a $w$ value, at the cost of additional computation time required by the minimization procedure.

Depending on the application, there are other efficient ways of choosing the regularization parameter $w$. Rather than using the same $w$ across an entire image or using different adaptive values of $w$ for each pixel, one could segment the image into regions of similar colors, and use the same parameter value for each region. Alternatively, if a large number of similar objects are to be imaged, one can directly measure the reflectance of a small subset and empirically find the optimal $w$ to be used for other objects. Another option would be to determine the most stable $w$ value using repeated leave-one-out cross-validation with various illuminants. Ultimately, the user can determine a suitable method for choosing the regularization parameter based on their experience.

\subsection{Design of multi-illumination imaging}

The choice of illuminants is an integral part of our method that bears on the accuracy and spectral range of the reconstructions. Our results indicate that the specific spectra of illuminants do not impact reconstruction accuracy as long as they are linearly independent and collectively cover the entire range of relevant wavelengths. Therefore, practical considerations for choosing the set of illuminants may focus on the spectral coverage and availability. Brightness is also important because the illuminants should ideally produce comparable signal levels, allowing large signal-to-noise ratios while keeping the camera response within its linear regime.

To further restrain the degeneracy in the spectral reflectance reconstruction, it is possible to supplement the use of multiple illuminants with a set of optical filters placed in front of the camera lens. Our use of an RGB camera effectively serves as three internal filters that are automatically applied to each illuminant. To use external filters, it will be important to avoid mechanically disturbing the camera and the samples, because our method requires that the images of the same sample under different illuminants are perfectly aligned. It is for this reason that we chose to use multiple illuminants that can be switched electronically, rather than optical filters that must be switched mechanically.

For the choice of camera, it is known that RGB cameras have reduced spatial resolution relative to a monochrome camera with the same pixel count, due to the need for demosaicing. For commercial RGB cameras to be used, they must allow access to raw images, manual exposure control, and the option to disable automatic image-processing such as white balancing \cite{Stevens}. In addition, RGB cameras are generally optimized for the human-visible range and typically require modification to image in the UV or near infrared, as was done in this work. We used an RGB camera in this work to demonstrate the easy implementation of our method. For better results, monochrome scientific cameras will offer high effective spatial resolution, high bit depth, low noise, and a calibrated linear response over a specified intensity range \cite{Howell2006CCD}. They are also available with extended sensitivity into the UV or infrared range \cite{HamamatsuImageSensorGuide}.

\subsection{Sample properties}

The most prominent discrepancy in our experimental results was the disagreement between the reconstructed reflectance and hyperspectral measurements of the butterfly specimens' yellow spots. We found that this was a consequence of the semi-transparency of the butterfly wings, as well as the bright lighting in the photo-box. Certain families of butterflies are known to have translucent or transparent wings \cite{butterfly_translucence, butterfly_wing_transparency}. In such cases, both reflectance and transmission may contribute substantially to the signal perceived by observers. Full characterization of a specimen's coloration may therefore require measurement of not only the specimen's reflectance and transmittance, but also the light spectrum of the surrounding environment \cite{Wainwright2025}. It is possible to extend our method to include transmittance reconstruction by modifying the imaging setup to enable both front and back illumination, rather than the near-uniform illumination of the current setup. Front illumination on a black background would enable reflectance measurement, while back illumination on a white background would enable transmittance reconstruction.

Many organisms achieve coloration through physical mechanisms distinct from absorbance, such as fluorescence and iridescence (structural color) \cite{Marshall2017Fluorescence, Glover2010StructuralColourPlants}. Fluorescence, wherein light of one wavelength is absorbed and re-emitted at a longer wavelength, could be included in a generalized method that includes a mathematical model of fluorescence, such as a low-rank Donaldson matrix \cite{Blasinski2020Fluorescence}. Its measurement requires additional experimental elements, such as the use of UV illumination in conjunction with UV blocking filters to isolate the emitted visible signal. It is also possible to extend our method to include iridescence and other forms of structural color, where the reflected light depends on both the angle of illumination and the viewing angle. This would require a more complicated setup similar to a gonioreflectometer, which allows full characterization of the bidirectional reflectance distribution function \cite{BRDF_source}.

\subsection{Metric for reconstruction error}

Because visual sensitivity generally falls off near the edges of an observer's spectral range, edge errors (such as those in \Cref{fig:Munsell}b) may have little effect on the predicted visual response. In such cases, spectral error metrics such as relative root mean square error (RRMSE), root mean square error (RMSE), and mean absolute error (MAE) can overstate the apparent severity of the reconstruction error. The reduced spectral agreement near the edges of the wavelength range is partly a consequence of weaker constraints in those wavelengths. Such inaccuracies have been noted in previous work using similar optimization methods \cite{Burns}. To address this issue, it is possible to define a narrower ``trusted'' spectral range after reconstruction. For example, although our reconstructions span 370--720~nm, the most reliable range is generally closer to 400--700~nm.

Conversely, reconstructions that appear highly accurate according to spectral error metrics (such as the example in \Cref{fig:color-checker-summary}a) can still produce colors that are perceptually distinguishable from the true colors according to the CIEDE2000 color-difference metric. This is because small spectral discrepancies can be perceptually important when they occur in wavelengths strongly weighted by the illuminant spectrum and observer sensitivity. Therefore, spectral error metrics can also overstate reconstruction quality. Together, these examples indicate that reflectance reconstruction should be evaluated not only by spectral agreement with a measured reference, but also by the degree to which it preserves the observer- and illuminant-dependent quantities for downstream analysis.

A main motivation for our reconstruction method is to be able to project the estimated spectral reflectance onto the visual systems of different animals. For example, in butterflies and birds where UV coloration and UV sensitivity are common, their color perceptions, rather than human vision, are essential for understanding color-mediated biological interactions. Moreover, perceptual differences can be substantial even within the human-visible wavelength range, due to differences in spectral sensitivity. This is clearly illustrated by our comparison of the butterfly image as viewed by human, avian, and butterfly visual systems (\Cref{fig:b3_combined}c); even with UV contributions ignored, the predicted visual patterns are already markedly different. We note that, in addition to sensory responses, the perception of visual patterns also depends on downstream cognitive processes, which requires knowledge of species-appropriate perceptual models and discrimination thresholds.

\section{Conclusion}

We have demonstrated that our method for reconstructing spectral reflectance of a sample from multi-illumination image data provides a robust alternative to hyperspectral imaging. We tested our method with synthetic image data generated from publicly available spectral reflectance datasets, achieving high accuracy with a modest number of illuminants and a wide range of the regularization parameter. We also demonstrated the use of our method to construct hyperspectral images of samples from digital camera images taken in a multi-illumination photo-box. Strong agreement with direct measurements verified the feasibility and robustness of our method.

This method is based on a physical model of color sensing involving the illuminant's spectral irradiance, the object's spectral reflectance, and the observer's visual sensitivities. Our method circumvents the problem of metamerism by leveraging the use of multiple illuminants and the smoothness of natural reflectance spectra. We have validated through both simulations and experiments that the method works accurately and robustly over a wide range of colors. Thus, it can be applied to a variety of materials with smooth reflectance spectra. Application of the method is not limited by the particular choice of illuminants (LEDs or broadband light sources), the imaging system (RGB or monochrome cameras), or the spectral range (UV, visible, infrared, etc.), as long as the illuminants' irradiance spectra and the camera's sensitivity functions cover that range. 

Overall, our results show that multi-illumination spectral reflectance reconstruction can bridge a gap between common digital photography and specialized hyperspectral imaging. Because direct hyperspectral imaging is often too slow or costly for large datasets, a reconstruction-based approach enables high spectral accuracy with easy implementation. In doing so, it provides an effective approach for studying visual signals objectively, and is highly scalable for application to large sample sizes, such as museum collections. Furthermore, the method is computationally simple, fast, and parallelizable, enabling high-throughput processing of image data. Our approach therefore significantly expands the capability of digital photography and creates a tool with broad applications to biology and other disciplines studying color vision and color patterns.

Multi-illumination image datasets have become increasingly available \cite{Murmann_2019_ICCV, Smagina_2019, Liu_2023}. Our method can be applied to such datasets to create hyperspectral or observer-specific image data for downstream analyses, such as machine-learning applications. Rather than training models directly on standard RGB images, one could train them on reflectance data or observer-specific representations, allowing biologically relevant dimensions of variation to be emphasized explicitly. For example, it is possible to ask more refined questions about the evolution of mimicry and other forms of signaling through coloration. Beyond biology, our method may be applied to samples where color characterization is important, such as analyses of paint pigments in art collections, identification of invasive species at ports of entry, and medical imaging of pathological tissues.

\section*{Acknowledgements}

We thank Leo Law and Kevin Hoppe for helping with imaging, Jay Horton for helping with making and painting a prototype photo-box, Shannon Heard and the Physics Machine Shop for making the photo-box, Pete Axson and the Physics Electronics Shop for wiring and installing the LED controls, Akito Kawahara and David Plotkin for granting access to the McGuire Center for Lepidoptera and Biodiversity, and Andrei Sourakov and Deborah Matthews for providing the butterfly specimens. This work is partially supported by the Physics Department Schaffer Undergraduate Scholarship, the Center for Condensed Matter Sciences Undergraduate Fellowship, and the College of Liberal Arts and Sciences University Scholars Program Fellowship to R.H., and by the University of Florida Research Opportunity Seed Fund to B.X.

\bibliography{sources}

\clearpage

\appendix
\crefalias{section}{appendix}
\crefalias{subsection}{appendix}
\setcounter{equation}{0}
\renewcommand{\theequation}{S\arabic{equation}}
\setcounter{figure}{0}
\renewcommand{\thefigure}{S\arabic{figure}}

\section*{Supplementary Material} 

\begin{figure}[h]
    \centering
    \includegraphics[width=\textwidth]{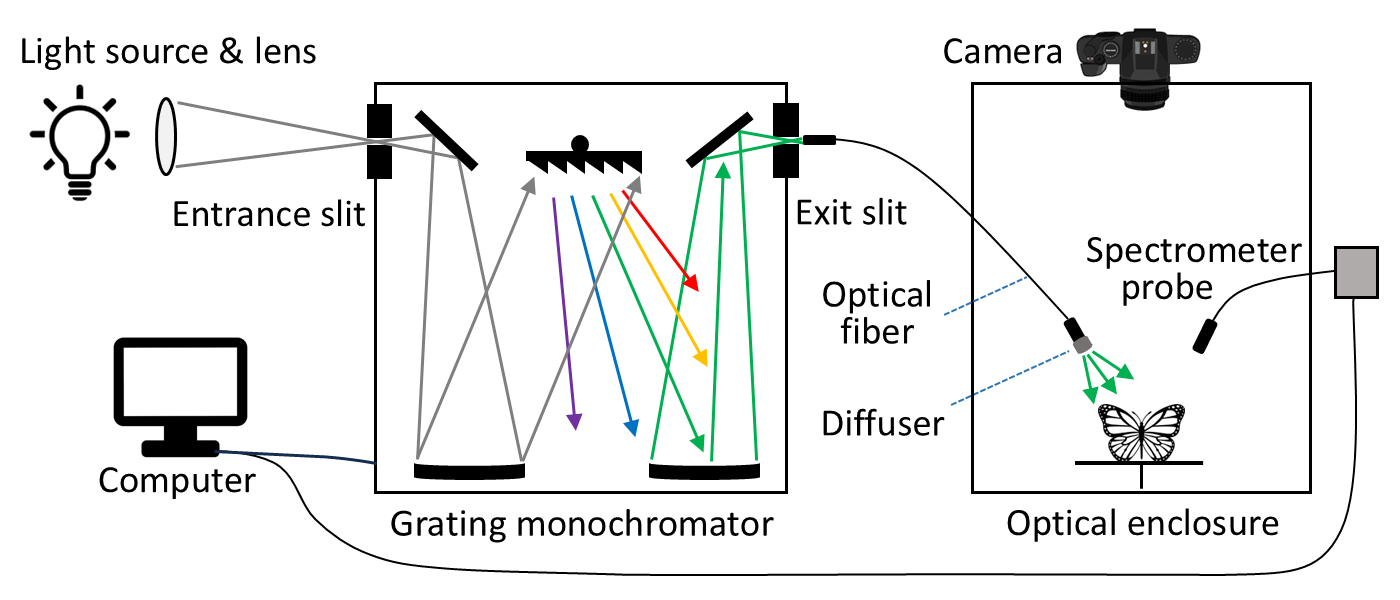}
    \caption{\small Diagram of the hyperspectral imaging setup. Light is focused onto the monochromator entrance slit. The internal grating is controlled by the computer to obtain a specific target wavelength output that is focused onto the exit slit, then passes into an optical fiber and through a diffuser onto the sample. The sample is then imaged with a monochrome camera to obtain an image at that target wavelength. The full set of images at each of the sampled wavelengths, referenced against images of a white standard, constitutes a hyperspectral image. The monochrome camera is replaced by the RGB camera to measure its sensitivity functions $c_i(\lambda)$; in this case, a white standard is used as the sample, and a spectrometer probe is used to simultaneously measure the intensity of the light reflected from the sample.}
    \label{fig:hyperspectral-setup}
\end{figure} 

\begin{figure}[t]
    \centering
    \includegraphics[width=\textwidth]{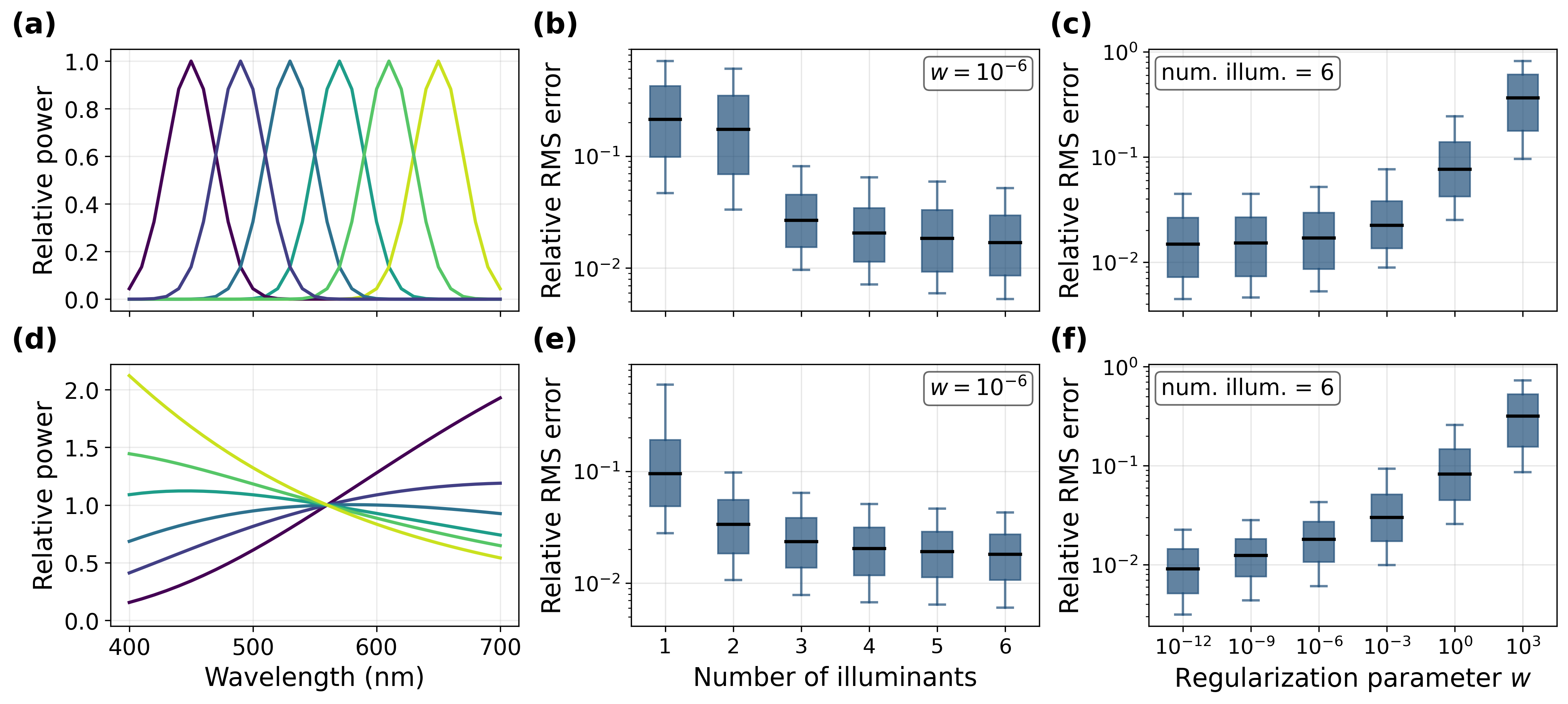}
    \caption{\small Robustness of the reconstruction method with respect to choices of illuminant spectra. The same procedure as in \Cref{sec:munsell} was used to generate CIE XYZ tristimulus values from the Munsell dataset. The sets of illuminants used are: (a--c) six evenly spaced Gaussian illuminants (idealizations of peaked LEDs) with means at wavelengths of $570, \, 650\, , 450, \, 530, \, 610, \, \text{and} \, 490\, \text{nm}$ (added in that order for (b)) and a standard deviation of $20 \, \text{nm}$; (d--f) six broadband blackbody illuminants at temperatures of $2900, \, 4000, \, 5000, \, 6500, \, 8000, \, \text{and} \, 12000$~K (added in that order for (e)). The blackbody illuminants are normalized to have a value of 1 at 560~nm, in accordance with the CIE convention for standard illuminants \cite{CIE_illuminant_convention}. (b,e) Box plots showing the relative root mean square error (RRMSE) statistics for reflectance reconstruction over the whole Munsell dataset as the number of illuminants is increased. The regularization parameter used is $w=10^{-6}$. (c,f) Box plots showing the RRMSE statistics as the regularization parameter is varied, with all six illuminants. Black lines indicate the median, blue boxes span the interquartile range, and and whiskers extend from the 10th to the 90th percentile.}  
    \label{fig:munsell-other-illuminants}
\end{figure}

\begin{figure}[t]
    \centering
    \includegraphics[width=\textwidth]{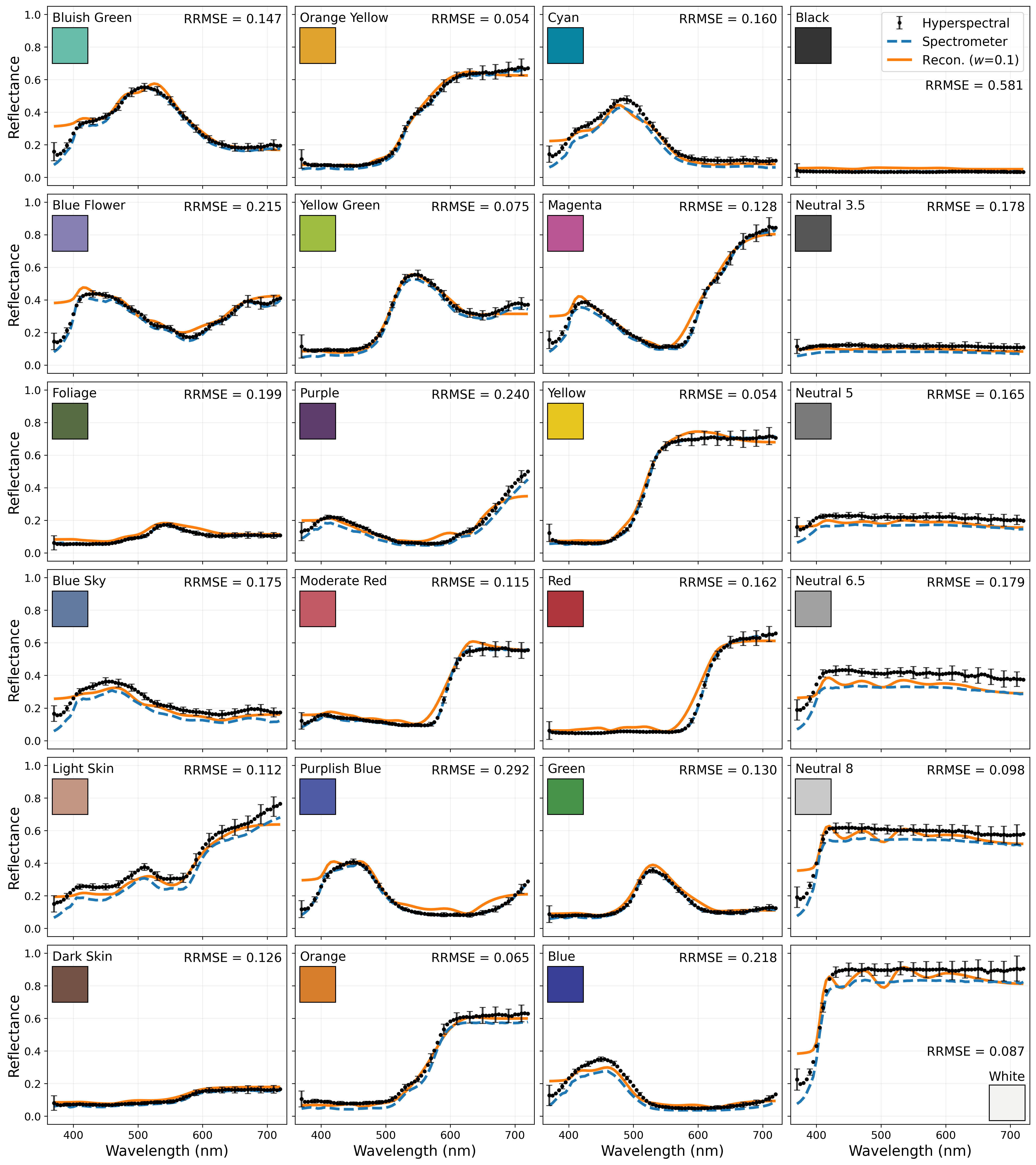}
    \caption{\small Spectral reflectance reconstruction of all 24 Calibrite ColorChecker patches. Orange curves show the reconstructed reflectance  using all eight LEDs and a regularization parameter $w=0.1$. Dotted black curves show the reflectance as measured with hyperspectral imaging; error bars indicate standard deviation, as calculated over a $10\times10$ pixel region in the hyperspectral image at each sampled wavelength. Dashed blue curves show the respective measurements using the spectrometer probe. Calculated RRMSE values compare the reconstructed spectra with the hyperspectral measurements. There is some discrepancy between the overall brightness of the measurements, especially for the gray scale patches whose reconstructed reflectance curves show spurious oscillations, which can be eliminated by using a larger value of $w$.}  
    \label{fig:color-checker}
\end{figure}

\begin{figure}[t]
    \centering
    \includegraphics[width=\textwidth]{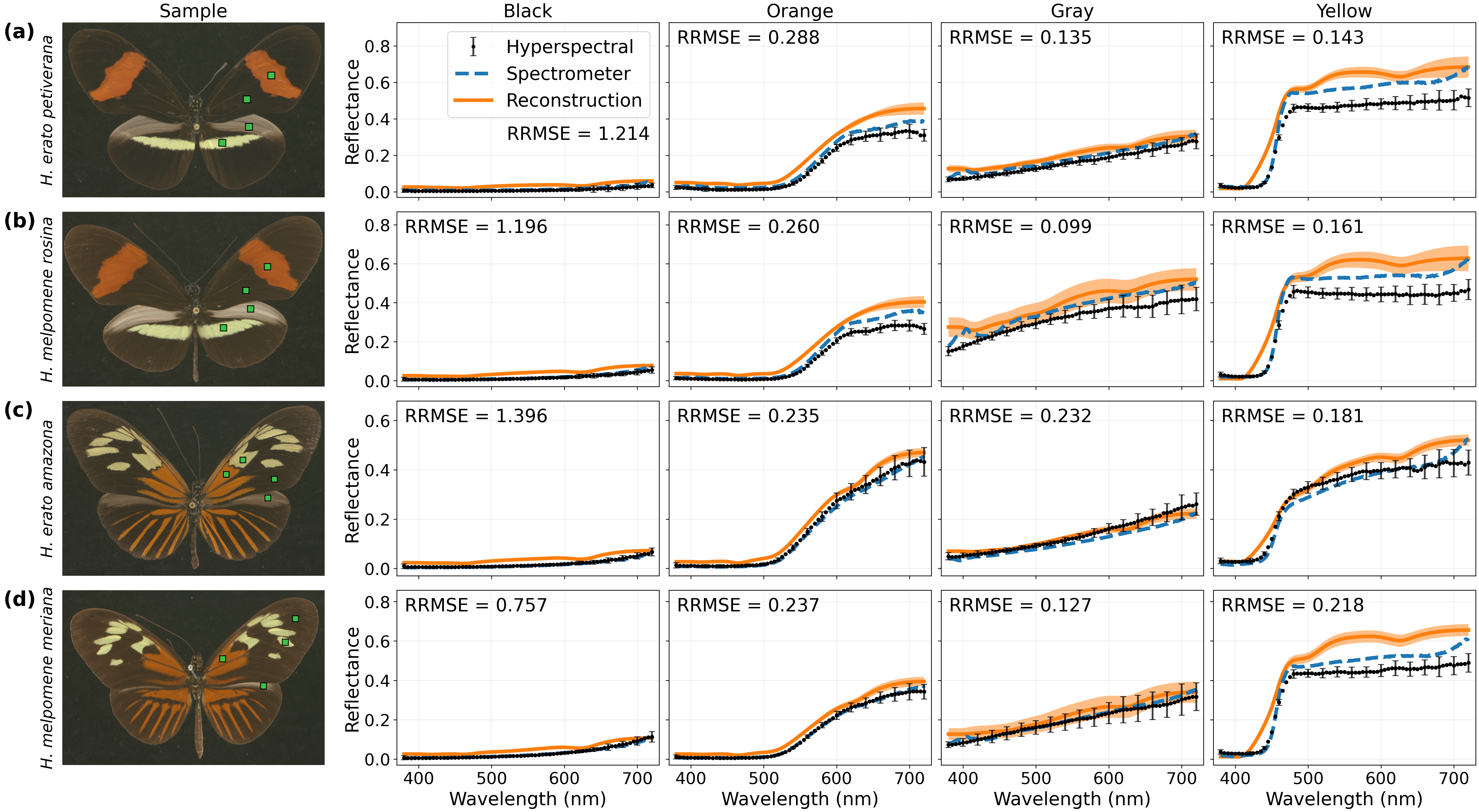}
    \caption{\small Spectral reflectance reconstruction of four selected color patches on the dorsal wings of four butterfly samples, using a regularization parameter $w=1$. (a) \textit{H. erato petiverana}, (b) \textit{H. melpomene rosina}, (c) \textit{H. erato amazona} (same as in \Cref{fig:b3_combined} of the main text), (d) \textit{H. melpomene meriana}. The left panels show the reconstructed image of each butterfly projected under daylight (D65) according to human color vision. The reflectance plots show the mean (solid orange curves) and standard deviation (light orange ribbons) over $10 \times 10$ pixel regions, which are indicated by the green squares in the left panels. Direct measurements using the hyperspectral imaging setup (black dots) and spectrometer probe (dashed blue curve) are shown as references. Error bars for the hyperspectral measurement represent standard deviation over a $10\times10$ pixel region. RRMSE values are calculated between the reconstructed reflectance and the spectrometer measurements. Systematic error in both shape and magnitude of the reconstructed reflectance (especially in the ``Yellow" patches) results from semi-transparency of the samples. The disagreement between the hyperspectral and spectrometer measurements is particularly stark, which is a consequence of the spectrometer measurements being performed in the high-illumination photo-box and the hyperspectral measurements being performed in the low-illumination hyperspectral imaging setup. The hyperspectral measurements are therefore expected to provide a more accurate measurement of wing reflectance.}  
    \label{fig:other-butterflies}
\end{figure} 

\begin{figure}[t]
    \centering
    \includegraphics[width=\textwidth]{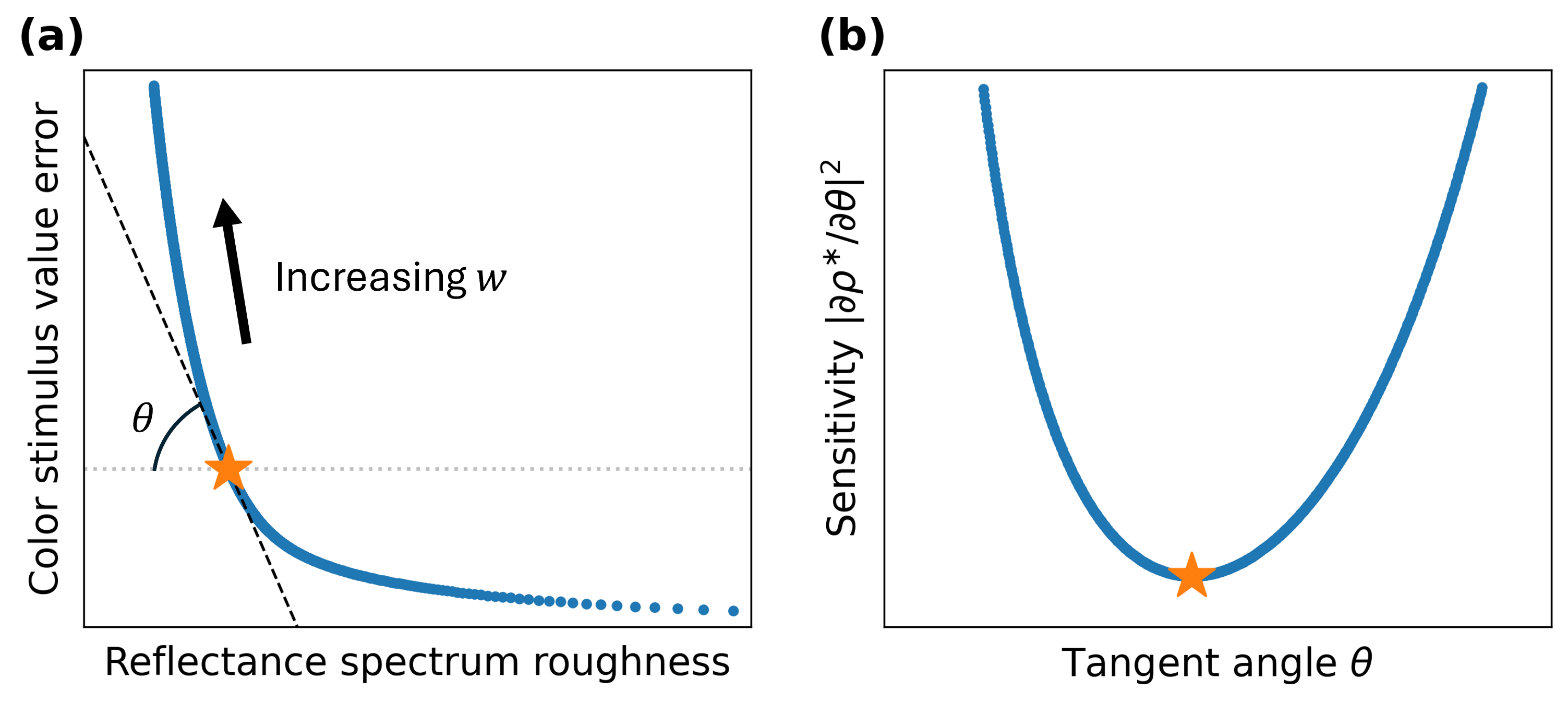}
    \caption{\small Optimization method for choosing the regularization parameter. (a) The Pareto curve shows the trade-off between the two terms in the objective function, which represent the color stimulus value error and the reflectance spectrum roughness. The parameter $w$ represents the negative slope along the curve, and each point on the curve represents a solution $\rho^*(w)$ for a set of color stimulus values at a single pixel. The dashed line is tangent to the Pareto curve and makes an angle $\theta = \arctan(w)$ with the horizontal. (b) A plot of $|\partial\rho^* / \partial\theta|^2$ versus $w$. The orange star represents the value of the regularization parameter $w$ where the reconstructed reflectance spectrum is most stable with respect to changes in $\theta$.}
    \label{fig:choosing-hyperparameter}
\end{figure} 

\end{document}